\documentclass[manuscript]{acmart}

\AtBeginDocument{%
  \providecommand\BibTeX{{%
    \normalfont B\kern-0.5em{\scshape i\kern-0.25em b}\kern-0.8em\TeX}}}

\setcopyright{acmcopyright}
\copyrightyear{2022}
\acmYear{2022}
\setcopyright{acmcopyright}\acmConference[FAccT '22]{2022 ACM Conference on Fairness, Accountability, and Transparency}{June 21--24, 2022}{Seoul, Republic of Korea}
\acmBooktitle{2022 ACM Conference on Fairness, Accountability, and Transparency (FAccT '22), June 21--24, 2022, Seoul, Republic of Korea}
\acmPrice{15.00}
\acmDOI{10.1145/3531146.3533182}
\acmISBN{978-1-4503-9352-2/22/06}

\begin{document}

\title{Designing for Responsible Trust in AI Systems: A Communication Perspective}


\author{Q. Vera Liao}
\affiliation{%
  \institution{Microsoft Research}
  \city{Montreal}
  \country{Canada}}
\email{veraliao@microsoft.com}

\author{S. Shyam Sundar}
\affiliation{%
  \institution{Penn State University}
  \city{State College}
  \country{USA}}
\email{sss12@psu.edu}

\renewcommand{\shortauthors}{}

\begin{abstract}
Current literature and public discourse on ``trust in AI'' are often focused on the principles underlying trustworthy AI, with insufficient attention paid to how people develop trust. Given that AI systems differ in their level of trustworthiness, two open questions come to the fore: how should AI trustworthiness be responsibly communicated to ensure appropriate and equitable trust judgments by different users, and how can we protect users from deceptive attempts to earn their trust? We draw from communication theories and literature on trust in technologies to develop a conceptual model called MATCH, which describes how trustworthiness is communicated in AI systems through \textit{trustworthiness cues} and how those cues are processed by people to make trust judgments. Besides AI-generated content, we highlight \textit{transparency} and \textit{interaction} as AI systems' affordances that present a wide range of trustworthiness cues to users. By bringing to light the variety of users' cognitive processes to make trust judgments and their potential limitations, we urge technology creators to make conscious decisions in choosing reliable trustworthiness cues for target users and, as an industry, to regulate this space and prevent malicious use. Towards these goals, we define the concepts of \textit{warranted trustworthiness cues} and \textit{expensive trustworthiness cues}, and propose a checklist of requirements to help technology creators identify appropriate cues to use.  We present a hypothetical use case to illustrate how practitioners can use MATCH to design AI systems responsibly, and discuss future directions for research and industry efforts aimed at promoting responsible trust in AI.


\end{abstract}






\maketitle

\section{Introduction}

With the popularity of complex AI systems used to augment or automate tasks that can affect many people's lives and have a long-lasting impact, trust is often cited as a key requirement for people to adopt AI technologies. The current academic and public discourses are predominantly structured around the guiding principles towards trustworthy AI~\cite{varshney2019trustworthy,toreini2020relationship},  often as a way to operationalize principles for responsible and ethical AI~\cite{mittelstadt2019principles}, such as ensuring effectiveness, fairness, transparency, robustness, privacy, security, and serving human values. These principles are inherently techno-centric, focusing on what constitutes the trustworthiness of AI, when in fact trust is a human judgment or attitude, which can be formally defined as a judgment of dependability in situations characterized by vulnerability~\cite{lee2004trust,vereschak2021evaluate}. The same AI technology can be judged differently by different people, with some forming inaccurate trust judgments. It is ultimately this psychological reality that determines how people would use and interact with the AI, and whether one could be harmed by inappropriate trust and consequent behavioral outcomes such as over-reliance and misuse.  

We argue that the AI field's fixation on trustworthiness results in blind spots in how people make trust judgments as well as how to \textit{communicate} the trustworthiness of AI appropriately and responsibly. Trustworthiness of a technology is not inherently established but communicated through \textit{trustworthiness cues}, which are embedded in interface features, documentation, and other modes of information, such as speech acts for conversational AI. With this communication perspective, our focus is on \textit{AI systems} rather than standalone models, where technology creators---including system developers and designers---need to make conscious decisions in choosing and designing these trustworthiness cues. The space of AI trustworthiness cues is becoming increasingly rich as researchers and practitioners build AI systems in numerous domains. The ethical imperative of transparency, in particular, calls for diverse types of information to be provided about the model's capabilities, limitations, decision processes, provenance, and so on. For example, the surging field of explainable AI (XAI) has produced a vast collection of techniques to generate model explanations~\cite{lipton2018mythos,adadi2018peeking,guidotti2018survey,doshi2017towards} with one goal, among others, being engendering trust in users.

By framing the design of AI systems as conveying trustworthiness cues, we foreground two issues that are of particular importance for holding AI technologies and their creators accountable. One is that malicious manipulation of trustworthiness cues can lead to undeserved trust with far-reaching harmful consequences. It is imperative to decouple the underlying model trustworthiness and the communication of it as a foundation to begin considering how to regulate AI system design. The other issue is that even well-intentioned technology creators may produce ill-designed trustworthiness cues that harm users due to a lack of attention to users' cognitive basis for trust judgments. The field's fixation on AI's trustworthiness can foster a false assumption that there are only ``ideal users'' who can perfectly assess it from available information. In reality, people have varied abilities and motivations to make accurate trust judgments. For example, abundant empirical evidence suggests that even technically sound AI explanations can result in harmful over-trust and over-reliance~\cite{zhang2020effect,bansal2021does,kaur2020interpreting,eiband2019impact,szymanski2021visual}. Some user groups are more vulnerable to these harms than others, such as AI novices~\cite{szymanski2021visual}, people working in cognitively constrained settings~\cite{robertson2021wait}, and even those with certain personality traits~\cite{ghai2021explainable}. We urge the AI field to develop a deeper understanding of how people process information to make trust judgments in order to develop reliable trustworthiness cues, as well as accountable mechanisms to generate them, all geared toward ensuring appropriate user trust and equitable user experiences with AI systems.  

To facilitate such an understanding, we present a conceptual model, named MATCH, focusing on the communication of trustworthiness in AI systems and the processes by which users make trust judgments (Section 3), by drawing from communication and human factors literatures (Section 2). MATCH decouples the trustworthiness attributes of the underlying AI model(s) and trustworthiness cues presented to the users, via three types of \textit{affordances} of AI systems: AI-generated content (e.g. predictions), transparency, and interaction. With this conceptualization, we highlight transparency as an affordance to enable trust judgments rather than warranting trust in itself, and bring forward the role of interaction design in shaping user trust. MATCH also highlights the wide variety of people's cognitive processes to make trust judgments: instead of always being processed analytically to form rational trust judgments, trustworthiness cues often invoke \textit{heuristics}---mental short-cuts or rules-of-thumb---for people to make speedy, but sometimes flawed, judgments. Communication theories further inform the varied tendencies to engage in heuristic trust judgment among different user groups.

On the basis of MATCH, we describe ``good'' trustworthiness cues that technology creators should use (Section 3.4). We define the concept of \textit{warranted trustworthiness cues} with a checklist of requirements to urge technology creators to focus on the psychological reality of their target users rather than technological qualities alone. We further suggest the use of \textit{expensive trustworthiness cues} as an industry practice that, by imposing a level of expense on technology creators, can help collectively guard against malicious means of deceiving user trust. 





To illustrate the use of MATCH, we present a case study of designing a hypothetical AI system (Section 4). While it is still a nascent area, we survey related works in human-computer and human-AI interaction to suggest a list of trustworthiness cues and trust heuristics. In Section 5, we reflect on the implications of MATCH and propose three areas of call to action to build responsible trust in AI: to regulate technology creators' use of trustworthiness cues, to empower users to make accurate trust judgments; and to look beyond model intrinsic attributes, and leverage social, organizational, and industrial mechanisms to enable reliable trustworthiness cues in AI systems.

\section{Background and related work}

\subsection{Trustworthy AI and trust in AI}
Ensuring the trustworthiness of AI, i.e., what's required for people to trust AI~\cite{varshney2019trustworthy}, has been considered an operational point to implement ethical AI principles~\cite{toreini2020relationship}. Building on the classic ABI (Ability, Benevolence, and Integrity) framework from the social sciences~\cite{mayer1995integrative,dietz2006measuring}, which prescribes trustworthy characteristics of a trustee as the three ABI dimensions,  \citet{toreini2020relationship} propose four categories of trustworthiness technologies for AI, namely Fairness, Explainability, Auditability and Safety (FEAS). The authors further demonstrate that these trustworthiness technologies operationalize core dimensions in many existing ethical and principled AI policy frameworks from both industry and governments. In a similar vein,  ~\citet{varshney2019trustworthy} maps out the trustworthy qualities of AI as predictive accuracy, robustness (to data shift and poisoning), fairness, interpretability, system-level provenance and transparency, and intention for social good. 

Another relevant thread of work explores organizational and regulatory ecosystems for ensuring trustworthy AI. \citet{shneiderman2020bridging} proposes a three-layer governance structure: reliable systems, safety culture, and trustworthiness certification by independent oversight.  \citet{knowles2021sanction} contend that the public distrust of AI originates from the underdevelopment of a regulatory ecosystem that would guarantee AI's trustworthiness. Drawing from the literature on institutional trust~\cite{anthony1984constitution}, the authors develop a model for public trust in AI-as-an-institution and highlight the pivotal role of auditable AI documentation in promoting public trust by constructing signals of trustworthy AI, establishing norms about what constitutes legal or ethical non-compliance, and allowing the exercise of control. 

Despite outlining the complexity of trust~\cite{siau2018building,toreini2020relationship}, these works are detached from the cognitive mechanism of how people make trust judgments. By examining the consequences of a collaborative system for data scientists, ~\citet{thornton2021fifty} demonstrate the nuances in implementing trustworthiness principles and highlight the gaps between these principles and actually promoting user trust. The latter requires attending to the designed aspects of the system that ``provide access to evidence of (dis)trustworthiness specific to a user, a technology and their context,'' or what they termed ``trust affordances.'' More recently, inspired by the literature on interpersonal trust~\cite{misztal2013trust}, ~\citet{jacovi2021formalizing} formalized human trust in AI as ``contractual trust,'' such that trust between a user and an AI model is anticipating that some contract will hold. Under this formalization, AI principles such as fairness, accountability, robustness, intention for social good, and privacy, can be seen as contracts, each of which places different criteria for people to establish trust, working together to form overall trust. This formalization brings forward the concept of \textit{warranted trust} (there exists a causal relationship between users' trust and the model's trustworthiness for a given contract). Accordingly, the authors suggest that the existence and level of warranted trust can be evaluated by manipulationist causality, i.e. whether and how much users vary their trust based on manipulated changes in the trustworthiness attribute of the model. 

\vspace{-1em}
\subsection{Trust in technologies: Lessons from communication and human factors literature}

Aligning with ~\citet{thornton2021fifty} and ~\citet{jacovi2021formalizing}, we aim to disentangle the relation between human trust and model trustworthiness. We further delve into cognitive aspects of trust judgments, for which we draw inspiration from the literature on trust in automation and web technologies. The two areas share emphases on considering the basis of trust, users' cognitive process to make trust judgments, and the impact of contextual factors, but they have different foci. Research on automation often studies human trust in association with the outcome of machine reliance, and dedicates much effort on elucidating the basis of trustworthiness, which we can draw parallels with the current emphasis on trustworthy principles of AI. Web trust literature deals with how people judge information quality and dependability on web sites~\cite{tseng1999credibility}, with the bulk of research conducted under the term ``web credibility''~\cite{rieh2007credibility}. For simplicity, we use the term "web trust" throughout the paper. 




\textbf{Trust in automation}. Our perspective is most directly informed by the seminal paper by~\citet{lee2004trust}. By synthesizing related literature, the paper proposes a conceptual model describing the process of trust formation in automation. Below we highlight a few key points of this model. 

\textit{Trust is determined by people's perception of information about the trustworthiness attributes of the system and existing beliefs.} There has been substantial work on conceptualizing trustworthiness attributes of automaton, which is often built on the ABI model~\cite{mayer1995integrative} mentioned above. \citet{lee1992trust} adapt the ABI model to three dimensions that more suitably characterize automated systems: performance (ability)---\textit{what} automation does;  process (integrity)---\textit{how} automation operates; and purpose (benevolence)---\textit why automation was developed. \citet{lee2004trust} show that many necessary characteristics of trustees discussed in the trust literature can be mapped onto these dimensions.

\textit{People's perception of trustworthiness attributes is mediated by the display of automation information, which is assimilated by multiple cognitive processes:} analytic, analogical (linking to known categories associated with trustworthiness), and affective (emotional response) processing. This perspective of multi-channel processing is key to understanding how people form trust judgments with rich displays that invoke trust-related heuristics and emotional responses.

\textit{Trust guides the behavior of reliance, but in a non-linear way, subject to the influence of individual, organizational and cultural contexts.} Several important factors influence the behavior of reliance, such as workload, intention for exploratory behavior, efforts to engage, perceived risk, self-confidence, time constraints, and system configuration. The dynamic interplay between automation, trust, and reliance can generate substantial non-linear processes: e.g., information display shapes the formation of trust, but existing level of trust also affects the selection and interpretation of information. Contextual factors also impact the development of trust directly. Individual differences vary the propensity to trust as well as channels of cognitive processing. Organizational and cultural contexts (e.g., other people's comments) play significant roles in trust development, highlighting the often neglected ecological aspects of trust.

\textit{Trust and appropriateness of trust are multi-faceted.} The scope of trust information display depends on the locus of trust, whether it is about trust in the system, function, or sub-functions. In the web trust literature, the locus of trust is differentiated for web, web sites, and messages~\cite{rieh2007credibility}. However, users are not always able to disentangle information and beliefs about them. Technology design should strive for appropriate user trust, which also has multiple facets: calibration (trust matches system trustworthiness), resolution (changes in system trustworthiness match changes in user trust), and specificity (able to differentiate different functional components of system trustworthiness). 


We also emphasize the mediating role of information display on trust judgments, and that appropriate trust relies on effective communication of system trustworthiness. To further elucidate the communication aspect, we now turn to the literature on trust in web technologies, which pays great attention to how interface cues shape trust judgments.

\textbf{Effects of information cues on trust in web technologies: communication perspectives and heuristic approaches}. Trust or credibility of information has long been studied in the fields of HCI, communication, and information science ~\cite{rieh2007credibility}. The early 2000s saw a rise in research on how people make trust judgments of web sites, including frameworks on what elements of web technologies influence people's trust~\cite{wathen2002believe,hilligoss2008developing,fogg2003prominence,sundar2008main,metzger2007making,flanagin2007role}. Practical means, including design guidelines~\cite{fogg2001makes,fogg2003users} and tooling~\cite{schwarz2011augmenting,yamamoto2011enhancing}, have also come out of this line of work both to facilitate technology creators' design choices and empower web users to make better judgments.

Much of this literature is based on communication theories of \textbf{dual-processes models} for attitude formation, including \citet{petty1986elaboration}’s elaboration likelihood model (ELM) and \citet{chaiken1980heuristic}’s heuristic-systematic model (HSM) (also related to \citet{kahneman2011thinking}'s System 1 and System 2 thinking). These theories postulate that web users engage in two cognitive processes to assess a website: one is ``systematic'' processing by paying attention to information content and performing a rigorous evaluation. The other is ``heuristic'' processing by attending to \textit{cues} about the information quality provided by the interface. The cues then trigger \textit{heuristics} that allow quick and cognitively easy judgments. A website can be seen as having two parts: its information content, and a repository of cues extrinsic to the content but contributing to trust judgments (e.g., article source, URL links, ``likes''), also referred to as ``content cues'' and ``contextual cues''~\cite{wathen2002believe} respectively.  Modern web technologies provide an abundance of contextual cues. Furthermore, dual-process theories predict that when users lack an ideal level of \textit{motivation} and \textit{ability} (broadly defined) to engage in systematic processing, they are likely to resort to heuristic judgments, often based on contextual cues.





This cue-heuristic perspective raises an important question: \textit{what cues are made available and what heuristics can be triggered by a given technology?} Web researchers have answered the question empirically~\cite{sillence2004trust,fogg2001makes}. By surveying 2500 participants, \citet{fogg2003prominence} summarizes 18 types of cues people frequently notice on a website to base their trust judgments on, such as information structure, name recognition, and advertising, with the most frequently mentioned cue being the ``design look.'' By conducting focus groups with 109 participants, \citet{metzger2010social} show that people routinely invoke cognitive heuristics to assess the trustworthiness of web sites, such as heuristics of reputation (e.g., website name recognition), endorsement (recommended by others or having good ratings), consistency (cross-validation in multiple websites), expectancy violation, and persuasive intent (e.g., advertising). 

Researchers also developed theoretical frameworks to account for the types of cues in web technologies~\cite{sundar2008main,wathen2002believe,hilligoss2008developing}. The MAIN model developed by~\citet{sundar2008main} has had a long-lasting impact. Its central thesis is that a given technology has certain ``affordances'' capable of cueing trust related cognitive heuristics (\textbf{affordance-cue-heuristic} approach). Affordance is an important concept in psychology and HCI (human-computer interaction) literature, defined as displayed properties of a system suggesting ways in which it could be acted upon or used~\cite{norman1988psychology,gibson1977theory}. The MAIN model earned its name by specifying four types of affordances that provide trust-related cues : Modality (e.g., visual modality cues realism heuristic), Agency (e.g., endorsement of ``other users'' cues bandwagon heuristic), Interactivity (e.g., the ability to customize cues control heuristic), and Navigability (e.g., the availability of many hyperlinks cues elaboration heuristic). Sundar summarizes a total of 29 heuristics that can be cued by these affordances~\cite{sundar2008main}. These cues can risk invoking unwarranted trust, as not all of them are directly linked to the content's trustworthiness, especially in malicious websites. Yet these cues play a major role in shaping trust judgments, given the deluge of online information and the impossibility of close scrutiny given cognitive limitations. This is true for all users, but especially for users who lack the ability or motivation to engage in a careful reading of the contents~\cite{sundar2008main}.



Equipped with these theoretical bases, we develop a conceptual model to describe trust judgments of AI systems in the next section. Our model synthesizes the above perspectives on the basis of trustworthiness, mediating role of information cues on trust judgments, the dual-process models, and the affordance-cue-heuristic approach.

\section{MATCH: a conceptual model of user trust in AI}
\begin{figure*}
  \centering
  \includegraphics[width=1\columnwidth]{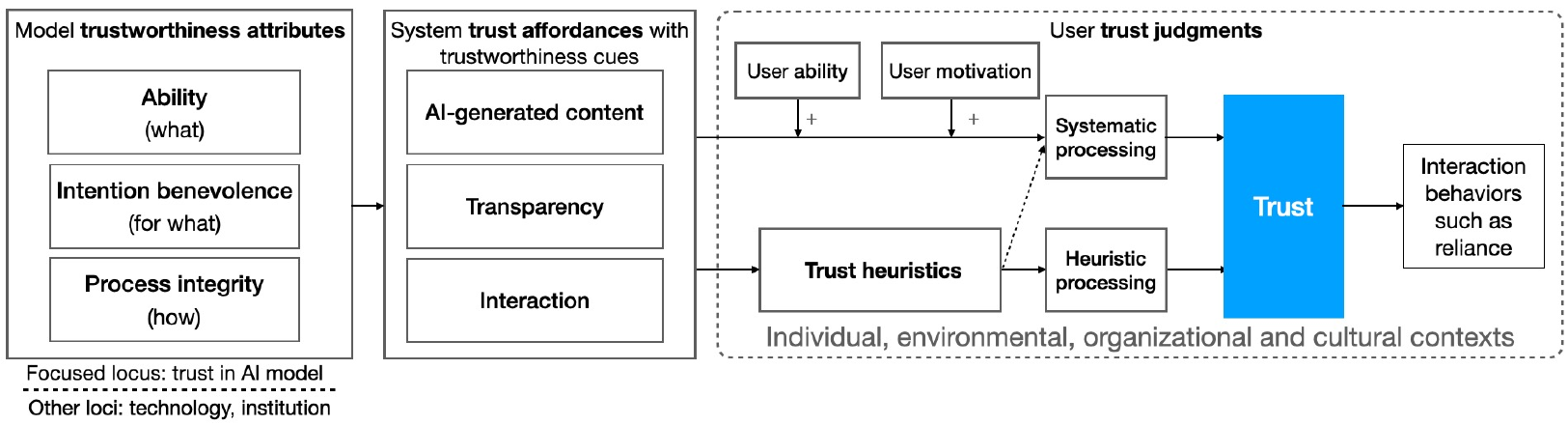}
    \vspace{-3em}

  \caption{MATCH model}~\label{fig:framework}
    \vspace{-3em}
\end{figure*}

Our conceptual model aims to describe how the trustworthiness of AI is communicated in AI systems and processed by users to make trust judgments (Figure~\ref{fig:framework}). The process is broken down into three parts: the underlying model (\textbf{M}) attributes, system affordances (\textbf{A}) to communicate trustworthiness (\textbf{T}) cues (\textbf{C}) of the AI, and users' cognitive processing of these cues by invoking trust-related heuristics (\textbf{H}). We refer to this model as MATCH and discuss each part below. 

As pointed out in Figure~\ref{fig:framework}, our scope is concerned with \textit{trust in the AI model(s)} that underlies a system, which we isolate from other loci of trust, such as trust in the institution (e.g., the company or brand producing the system)~\cite{rieh2007credibility} and trust in AI-as-a-technology~\cite{knowles2021sanction}. We focus on trust as an attitude rather than its effects on user behaviors such as reliance, and acknowledge that there are ecological factors beyond our focus on the internal cognitive processes that shape trust judgments and reliance, depending on individual, environmental, organizational, and cultural contexts.
\vspace{-0.5em}

\subsection{Model trustworthiness attributes: what makes the basis of trustworthiness? }

As reviewed, many define the trustworthiness of technologies~\cite{lee1992trust,lee2004trust,toreini2020relationship,kunkel2019let} based on the classic ABI model. \citet{lee1992trust} adapted ABI to the context of automation using the dimensions of performance, process, and purpose.  We adopt these dimensions and define the basis of AI trustworthiness as ability, intention benevolence, and process integrity. Note that the three core components of MATCH (trustworthiness attributes, trust affordances, and cognitive processes of trust judgments) should be agnostic to how the attributes are operationalized. We welcome future work to expand these dimensions or explore alternative operationalizations. In Figure~\ref{fig:prcess}, we suggest a non-exhaustive list of example attributes under each dimension, which will be discussed with an illustrative use case in Section 4.

For the sake of simplicity, our discussions in this paper assume that there is a single underlying model that serves as the basis of user trust in the AI system. For complex systems with multiple models or technological components, these dimensions would hold but one may need to define specific attributes differently (e.g., modular and joint abilities) and consider trust \textit{specificity} for different components as one facet of appropriate trust~\cite{lee2004trust}. 

 \textbf{Ability} refers to the capabilities of the underlying AI model with regard to its output or the function it provides to the user (e.g., making predictions, generating answers). They cover \textit{what} the AI can do. For example, \textit{overall performance} is a key attribute of AI ability.  Considering other trustworthy AI principles~\cite{toreini2020relationship,varshney2019trustworthy}, \textit{performance fairness} (e.g., absence of performance differences between different protected demographic groups) and \textit{performance robustness} (e.g., against data shift and poisoning~\cite{varshney2019trustworthy}) are also ability attributes. We also consider \textit{improvability} as an ability attribute. For example, an active learning model is set up to be improvable with user input. Note that this conceptualization distinguishes between objective performance as an underlying attribute of the model and performance metrics as trustworthiness cues to approximate (e.g., calculated using test data) and communicate the attribute.

\textbf{Intention benevolence} refers to the degree of benevolence behind the creation of the technology-- \textit{for what} is the AI developed? Besides \textit{intended use} (e.g., social good, serving user values), we also consider \textit{intended compliance} (e.g., privacy-preserving, security conscious) as an attribute of intention benevolence. 
    
\textbf{Process integrity} is the degree to which the operational or decision process of the model is appropriate to achieve the users' goal, describing \textit{how} the AI works. The standard of integrity should be context- and user-dependent, such as the absence of flawed logic, optimizing for the right goal, and aligning with known processes in the domain. While the level of process integrity could make a difference in the AI's ability attributes, the former creates a different basis of trust, one that is focused on the system's dispositional integrity rather than its outcomes.

These three dimensions of attributes determine the level of trust that users should have in an ideal world. However, in reality, these attributes are communicated through trustworthiness cues, and then the cues are judged through a plurality of cognitive processes, both of which introduce noise, as we discuss in the following. 

\subsection{Affordances for trustworthiness cues: how is trustworthiness communicated in AI systems?}
A \textit{trustworthiness cue} is any information within a system that can cue, or contribute to, users' trust judgments. For individuals, trust is often conceived as a judgment of dependability, and not necessarily driven solely by cues that explicitly reflect the three bases of trustworthiness described above. An AI system can thus place its users in a rich environment of trustworthiness cues. According to the affordance-cue-heuristic approach~\cite{sundar2008main}, one may identify trustworthiness cues by conceptualizing the \textit{affordances} of a given type of technology that engender such cues. 

The question of ``what are the trust affordances~\cite{thornton2021fifty} of AI systems'' can be challenging to answer since AI is far from a monolithic set of technologies. We base our proposal on currently popular AI systems (e.g., decision support, task assistance, recommender systems) and common system features in production and literature. A recent survey paper~\cite{lai2021towards} maps out AI system elements that have been empirically studied for AI decision support in HCI and AI literature, including different types of prediction output, information about the prediction (e.g., local explanations, uncertainty information), information about the model (e.g., performance metrics, documentation, model-wide explanations, training data), and user control features (e.g., customization, feedback to improve the model). Accordingly, we suggest three types of common affordances of AI systems: AI-generated content, transparency, and interaction. We discuss these affordances below, and suggest a list of example trustworthiness cues provided by these affordances in Figure~\ref{fig:prcess}.

\textbf{AI generated content} refers to displays of the model output or the functional support provided by the AI system. Depending on the type of model, displays can take the form of a predicted class label, a score, a list of suggestions, generated texts or images, etc. These displays can serve as direct trustworthiness cues for users to assess the ability attributes of the AI model. In some cases the design, e.g., under what circumstances AI assistance is provided or not, can also cue users' judgment of the intention benevolence of the model.  
    
\textbf{Transparency} affordance refers to displays allowing a better understanding of the model, broadly defined, including its behaviors, processes, development, and so on. We single out transparency as a unique affordance of AI systems given the increasing industry emphasis on providing transparency, exemplified by the prevalence of normative metrics (including performance, fairness, and robustness metrics), XAI features, and model documentation~\cite{mitchell2019model,arnold2019factsheets,hind2020experiences} (commonly including provenance information~\cite{knowles2021sanction,thornton2021fifty} about how and for what it was developed). Recent literature also discusses governance structures to ensure trustworthy AI~\cite{shneiderman2020bridging,raji2020closing}, such as internal reviews, testing, independent and government oversight, and so on. Communicating the process and outcomes of such governance structures should also be considered a form of transparency. Transparency allows cueing for all three dimensions of trustworthiness attributes---ability (e.g., through metrics), intention benevolence (e.g., communicating intended use and compliance in the documentation), and process integrity  (XAI features). This conceptualization highlights the role of transparency as an affordance for users to base their trust judgments on, rather than inherently warranting trustworthiness. This is related to \citet{jacovi2021formalizing}'s formalization of the goal of XAI as facilitating appropriate trust by increasing the trust of users in a trustworthy AI system and distrust in a non-trustworthy one. This goal can only be attained if trustworthiness cues in the transparency affordance are both truthfully communicated and appropriately assessed by the user.
    
\textbf{Interaction} affordance refers to displays that suggest how users can interact with the system, beyond the content of the model output, for which we consider both perceptual affordances (e.g., medium and design look) and action affordances (e.g., customization of the system, socialization possibilities with other people). The roles of interaction and interaction design are often overlooked in the current literature on trust in AI. We draw parallels with the web trust literature showing that people base their trust judgments not only on ``content cues'' but also on many ``contextual cues''~\cite{petty1986elaboration} on a web site, such as the design look, source reputation, or social information~\cite{fogg2001makes,fogg2003prominence,metzger2010social}. Some interaction affordance is directly enabled by the model ability, such as customization in guiding the model's behavior, and can directly cue the trustworthiness attributes of ability (improvability) and intended use (e.g., serving user preferences). Other interaction affordances may be extrinsic, even irrelevant, to the model (e.g., the choice of medium, such as using a visualization), but can still cue people's trust judgments. By bringing to light interaction as an affordance providing rich trustworthiness cues, we urge future research to better understand whether and how different interaction features of AI systems, even disassociated from the underlying model, can impact user trust.

\vspace{-1em}

\subsection{Dual cognitive processes: how are trustworthiness cues processed by people?}

MATCH conceptualizes this process based on dual-process models of attitude formation~\cite{petty1986elaboration,chaiken1980heuristic}. The basic idea is that people process information to form a judgment or attitude through two routes: 1) \textbf{systematic processing} by rigorously assessing the information to make a rational judgment, and 2) \textbf{heuristic processing} by following known heuristics or rules-of-thumb to make a cognitively easier judgment. However, when the heuristics are applied inappropriately, the judgment is prone to errors. MATCH further highlights the roles of trust heuristics and individual differences.

\begin{figure*}
  \centering
  \includegraphics[width=0.9\columnwidth]{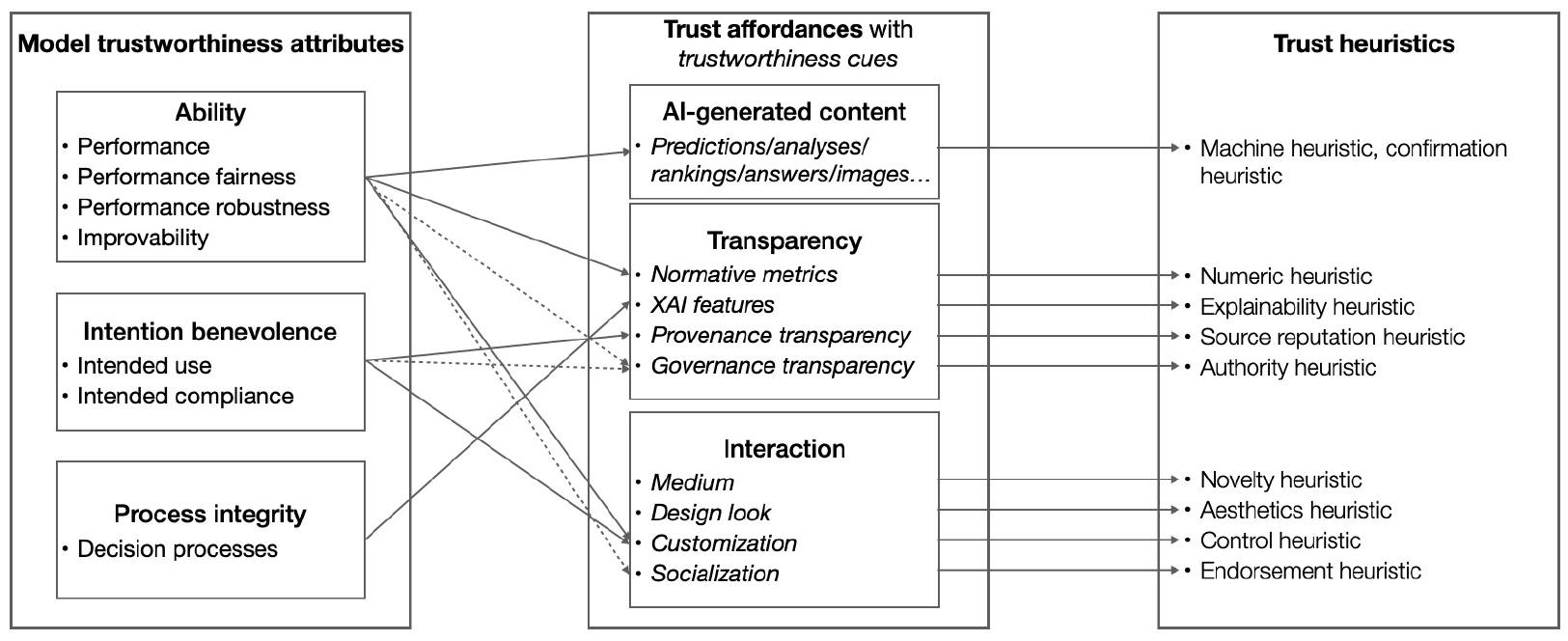}
  \vspace{-1em}

  \caption{Example lists of trustworthiness attributes, trustworthiness cues and heuristics, used in the use case in Section 4}~\label{fig:prcess}
    \vspace{-2.5em}
\end{figure*}

\textbf{Trust heuristics} are any rules of thumb applied by a user to associate a given cue with a judgment of trustworthiness. There are many ways for individuals to acquire trust heuristics. Some are common cognitive heuristics applied to the context of AI. For example, online users tend to apply the \textit{authority heuristic} by following the opinion of an authority on the subject matter~\cite{metzger2010social}. This heuristic can be invoked by an interface cue showing certification from a regulatory body that audited the AI. Others are technology-specific heuristics learned from past experiences. For example, some groups of users may have a prominent \textit{machine heuristic}, believing machines are more reliable than humans~\cite{sundar2019machine}. The phenomenon of XAI features leading to over-trust~\cite{zhang2020effect,bansal2021does,kaur2020interpreting,eiband2019impact,szymanski2021visual} can be attributed to an ``\textit{explainability heuristic}''~\cite{liao2021human,ehsan2021explainable} that superficially associates being explainable with being capable, without deliberating on the actual content of the explanation. Heuristics can also be intentionally cultivated by technology creators. One example is to provide instruction and supporting evidence that a number above a certain threshold of a normative metric could be considered acceptable. The existence of heuristics varies between individuals. However, it is possible to enlist common heuristics based on psychology and communication theories~\cite{sundar2008main,metzger2010social}, or by empirically exploring the heuristics that are frequently invoked during target user groups' interaction with the AI, by using, for example, think-aloud methods to examine cognitive processes~\cite{jaspers2004think,hoffman2018metrics}. It is worth pointing out that while heuristic processing necessarily involves heuristics, heuristics can also be used to aid systematic processing when they are applied in a conscious and deliberative fashion~\cite{sundar2008main,chaiken1980heuristic}. In Figure~\ref{fig:prcess}, we provide example heuristics based on prior literature, to be discussed in Section 4.

\textbf{Individual differences in systematic vs. heuristic processing}. People have different tendencies to engage in systematic versus heuristic processing. Hence, the introduction of a trustworthiness cue can risk creating inequality in trust and user experience. For example, recent studies have repeatedly found that XAI features bring less benefit, or even harms (leading to over-trust), to certain user groups such as AI novices or users working in cognitively constrained settings~\cite{szymanski2021visual,ghai2021explainable,robertson2021wait}. Research based on dual-process models~\cite{petty1986elaboration} has long established that when individuals lack either the \textit{motivation} or \textit{ability} to perform systematic processing and rationally assess trustworthiness, they are likely to resort to heuristics. Note that motivation and ability are umbrella terms that can encompass many user and contextual characteristics, which make these theoretical models powerful for understanding and predicting individual differences. For example, a user may lack ability due to a lack of AI knowledge, domain knowledge, or cognitive capacity; they may lack motivation due to perceived cost versus gain, personality traits, or competing motives~\cite{petty1986elaboration,petty1984source,o2013elaboration}. By highlighting these individual differences, we encourage technology creators to carefully examine and mitigate the potential inequalities of experience among users who may not have an ideal profile of ability or motivation.

\vspace{-0.5em}
\subsection{What are ``good'' trustworthiness cues?}
Based on this conceptual model, we attempt to address an important question: \textit{what are ``good'' trustworthiness cues that should be used by technology creators?} It is helpful to break down the consideration of ``goodness'' into two scenarios: 1) for a well-intended technology creator, a good trustworthiness cue is one that results in well-calibrated trust judgments by target users with regard to the true trustworthiness of the AI; 2) for the industry and society as a whole, a good trustworthiness cue is one that both has good calibration and is likely used truthfully to communicate the underlying trustworthiness, or in other words, not subject to malicious and deceptive use. 

\textbf{Warranted trustworthiness cue}. To facilitate efforts around using and regulating good trustworthiness cues, we first introduce this concept. We consider a trustworthiness cue to be warranted if:
\vspace{-0.5em}
\begin{enumerate}
    \item It is truthfully used by the technology creator, without deceptive intentions (\textbf{truthfulness condition}).
    \item It is relevant to or reflective of the underlying model trustworthiness attributes, including ability, intention benevolence, and process integrity (\textbf{relevance condition}).
    \item It leads to well-calibrated trust judgment by the target users with regard to the trustworthiness attribute(s) it reflects (\textbf{calibration condition}).
\end{enumerate}{}
\vspace{-0.3em}
This concept is relevant, but not identical to~\citet{jacovi2021formalizing}'s formalization of warranted trust (if the incurred trust corresponds to the trustworthiness of the model), as our focus is on what kind of trustworthiness cues are likely to result in warranted trust in users. We now discuss the requirements for achieving each of these conditions. 

\textbf{Relevance condition.} This condition underscores the need to consider if a trustworthiness cue is indeed reflective of the underlying trustworthiness attributes of the AI. Technology creators should pay attention to \textit{prominent irrelevant trustworthiness cues}---cues that shape users' trust judgments, but are not reflective of the trustworthiness attributes of the model, such as the surface design look or a link to an external web site. Often, some irrelevant cues are unavoidable because they support other user goals, but they can impact user trust in unintended ways. Such unintended effects can be mitigated by making these cues less prominent during users' trust-development stage, providing interventions to disrupt invocation of trust heuristics (e.g., a reminder that the design is inherited from a template), or guiding user attention to other relevant trustworthiness cues. 

While we encourage technology creators to incorporate comprehensive trustworthiness cues that directly describe the model trustworthiness attributes of ability, intention, and process, the relevance condition should embrace any cues that provide supporting evidence for these attributes. We may differentiate between \textit{model-intrinsic} and \textit{model-extrinsic trust-relevant cues}. Intrinsic cues are generated directly from the model or its development process, such as its output, performance metrics, and explanations. Extrinsic cues are generated from social, organizational, and industrial processes outside model development but can provide supporting evidence for its trustworthiness attributes. Examples may include other users' reviews, audit trails, and evidence from external or regulatory oversight.

\textbf{Calibration condition.} Calibration requires a match between a person's trust judgment based on a given trustworthiness cue and the true trustworthiness of the underlying model attribute(s) reflected by the cue. Calibration can be assessed with a \textbf{formal analysis}---measuring the change in people's trust judgment, by manipulating the quality of the corresponding model trustworthiness attribute(s) reflected by the trustworthiness cue ~\cite{jacovi2021formalizing}. A 
good trustworthiness cue should result in consistent changes between the two. Calibration should be considered probabilistic instead of a dichotomy---e.g., showing an accuracy metric may provide better calibration than showing predictions alone, but neither may be perfectly calibrated to the true changes in model ability.


However, it is not always feasible to perform costly formal analysis to quantify the calibration, which may also suffer from generalizability issues given individual and contextual differences. Based on MATCH, we suggest the following heuristics to help technology creators identify whether a given trustworthiness cue has a high or low probability of calibration. We postulate that a trustworthiness cue is more likely to satisfy the calibration condition if:
\begin{itemize}
    \item \textit{The target user group has the ability and motivation to perform systematic processing} (\textbf{systematic condition}).

\end{itemize}{}
 \vspace{-0.5em}
 
\textbf{Or}
 \vspace{-0.5em}
\begin{itemize}
    \item \textit{It does not invoke unfounded trust heuristics} (\textbf{no unfounded heuristic condition}).
\end{itemize}{}

Whether a heuristic is ``unfounded''---with little evidence to support or low probability to hold---depends on the context and the user. For example, a recent study~\cite{lu2021human} shows that people follow the cognitive heuristic of \textit{confirmation bias} when viewing AI predictions, whereby the agreement of AI predictions with their own judgment is seen as indicating high AI ability. This heuristic may be unfounded (low probability to hold) if users are novices to the decision task, but could be acceptable for domain-expert users. Meanwhile,  some heuristics are generally unfounded and we should avoid invoking them---for example,  recent studies of user interaction with XAI suggest the existence of unfounded \textit{explainability heuristic}~\cite{liao2021human}
(associating being explainable with being capable) and \textit{numeric heuristic} (associating numerical explanations with algorithmic intelligence). According to the prominence-interpretation theory on web trust~\cite{fogg2003prominence}, the existence of unfounded heuristics should best be assessed jointly by the likelihood of noticing a feature and invoking a trust heuristic (prominence), and the likelihood of the heuristic being unfounded (interpretation).

Importantly, this condition acknowledges the benefit of \textit{founded heuristics}. Heuristics are an indispensable part of people's cognitive process and it is unrealistic to expect all users to have the ability and motivation to perform systematic processing at all times. Individuals take advantage of heuristics because they are founded in past experiences or some conditions, and can generally help them make better judgments. Technology creators should strive to leverage common cognitive heuristics, reverse-engineer reliable mechanisms to make users' naturally invoked heuristics better founded, and cultivate founded heuristics by providing training, guidance, or reinforcement mechanisms. For example, it is known that people have a common \textit{anchoring heuristic/bias} that hinges judgment on their first encounter with an object of trust. A design choice that makes this heuristic better founded in the context of AI systems is to present users with performance transparency information during the system on-boarding stage. This kind of effort is key to engendering equitable trust by enabling users with less-than-optimal motivation and ability to better assess AI systems.

\textbf{Truthfulness condition.} We consider this condition last as it is concerned with the intent of technology creators. Rather than asking what is required for this condition, the key question here is how to prevent deceptive use of untruthful trustworthiness cues. We bring in one perspective by drawing on ``costly signaling theories'' from evolutionary psychology~\cite{zahavi1975mate}. Signaling theories are a body of theoretical work~\cite{connelly2011signaling,bliegebird2005signaling} concerned with how individuals (humans and animals) select signals (traits, actions, etc.) to present during communication to convey some desirable quality for achieving a social goal. Since individuals have motivations to deceive, collectively evolution would favor reliable signals that are ``costly''---costing the signaller something that could not be afforded by those with less of a given quality.

With a similar motivation to collectively guard against deception, we argue that the industry as a whole should prioritize using \textbf{expensive trustworthiness cues} that would impose a level of expense on technology creators. We consider ``expense'' as any investment that a creator must make to present a trustworthiness cue to a believable extent to the users, including but not limited to development, time, and infrastructure expenses. For example, showing an accuracy metric is less expensive than a user-friendly XAI visualization; establishing positive audit trails and endorsement from others are generally costly in terms of time and effort. More expensive trustworthiness cues also include comprehensive documentation, certification from established review boards, and customization features. However, in practice, individual technology creators may need to weigh the expenses and limit their choices to cues that are within their affordable range. Like many responsible AI practices, costly implementation runs the risk of marginalizing smaller business entities and creating inequalities in the industry. While we suggest leveraging expense on technology creators to safeguard the truthfulness condition, a much more nuanced view on its relations with resources, gains, and other motivators and constraints needs to be developed to inform policy and industry practices.



\section{Use case: using MATCH to design for appropriate trust in an AI symptom checker}

MATCH can help technology creators prospectively interrogate what trustworthiness cues should be presented in a system, or retrospectively understand the causes of users' inappropriate trust (e.g., whether due to salience of trust-irrelevant cues or miscalibration). We demonstrate the former with a hypothetical use case of designing an AI system. We also leverage this use case to present examples of model trustworthiness attributes, trustworthiness cues, and trust heuristics to help ground our conceptual model, as summarized in Figure~\ref{fig:prcess}. We start by describing the use case:
\begin{quote}
\vspace{-0.5em}
\textit{HealthChecker is an AI app that suggests diagnosis for common diseases based on a list of symptoms that a user provides. Its suggestions also take into consideration the patient's personal information, such as demographic and socioeconomic background, and health sensor data if the patient consents to their collection. HealthChecker also sends its suggestions to the patient's primary doctor for verification and suggestions. }

\textit{There are two groups of primary users. One is  patients, to represent which the creators consider the persona ``Eric''---an average mobile app user who has needs for diagnosis of common diseases a few times a year, is neither an AI expert nor health expert, but is keen on trying out new technologies. The other is the patients' primary doctors, for which the creators consider the persona ``Jessie''---an average primary care doctor who is a medical expert but only moderately familiar, and usually cautious, with AI technologies.  }
\vspace{-0.5em}
\end{quote}

\textbf{Step 1: Which model attributes determine its trustworthiness and should be communicated?} The creators start with this question. Based on MATCH, they consider the model's ability attributes including performance, fairness, robustness and improvability, intention benevolence that governed the model development, including intended use and compliance (e.g., privacy-preserving is especially important here), and the model's process integrity to make diagnoses.
 
\textbf{Step 2: What kinds of cues should be used to communicate trustworthiness?} This analysis happens in parallel with other design decisions for the system, such as the kind of interface needed to support efficient use and a user-friendly experience. MATCH guides them to consider what trustworthiness cues should be designed into the affordances of AI-generated content, transparency, and interaction. For AI-generated content, the diagnosis suggestion itself can cue judgments of the AI's ability. They start with a simple design of presenting only the top suggestion. 

For transparency, they consider multiple features that communicate the three dimensions of trustworthiness attributes: normative metrics for accuracy, fairness, and robustness to show ability; explanations of diagnoses to show process integrity; and documentation that highlights the intended use and compliance considerations of the AI. The solid arrows in Figure~\ref{fig:prcess} show the mapping analysis between trustworthiness cues and trustworthiness attributes. They also consider what kind of model-extrinsic cue can provide supporting evidence for the model's trustworthiness attributes. An internal review board has been introduced in the company to oversee the development of AI technologies, reviewing regulatory and ethical compliance such as fairness and privacy. The creators choose to include information about this governance structure and a certification from the board in the documentation. The dashed arrow in Figure~\ref{fig:prcess} shows the mapping between model-extrinsic cues and the trustworthiness attributes they support.

For interaction, the creators decide to invest in a customization function as an important trust-building feature, by allowing users to choose if they want to provide certain personal information. This feature lets users experience the improvability of the AI and the compliance intention to preserve privacy. The creators also need to examine other interaction features. The app has a built-in socialization feature that allows patients to see the doctors' ratings and feedback for diagnoses made by the app. The creators realize that patients will likely use such information to assess the AI's ability, as a type of model-extrinsic trustworthiness cue. By observing and talking to some users like Eric, the creators realize that the sleek design of the app and the use of a chatbot to gather information about their symptoms contributed significantly to their trust in the system. The design look and medium, however, cannot be mapped to the model trustworthiness attributes and should be considered irrelevant trustworthiness cues. 

\textbf{Step 3: Are these cues warranted trustworthiness cues for both user groups? If not, why?} This question can help foresee inappropriate and inequality of trust, and identify causes for improvement. It should be asked iteratively as the design progresses or whenever a new system feature is introduced. The creators leverage the conditions for warranted trustworthiness cues discussed in Section 3.4 to analyze the cues identified above. We assume that the truthfulness condition is satisfied for the creators of HealthChecker. For the relevance condition, the analysis above identifies that the medium (a chatbot interface) and the design look do not satisfy the condition, and can potentially trigger novelty and coolness heuristics~\cite{sundar2008main} that lead to positive trust judgments. If over-reliance occurs, interventions should be introduced to either tone down these cues for new users or mitigate the prominence of triggered heuristics.
 
\begin{figure*}
  \centering
  \includegraphics[width=0.7\columnwidth]{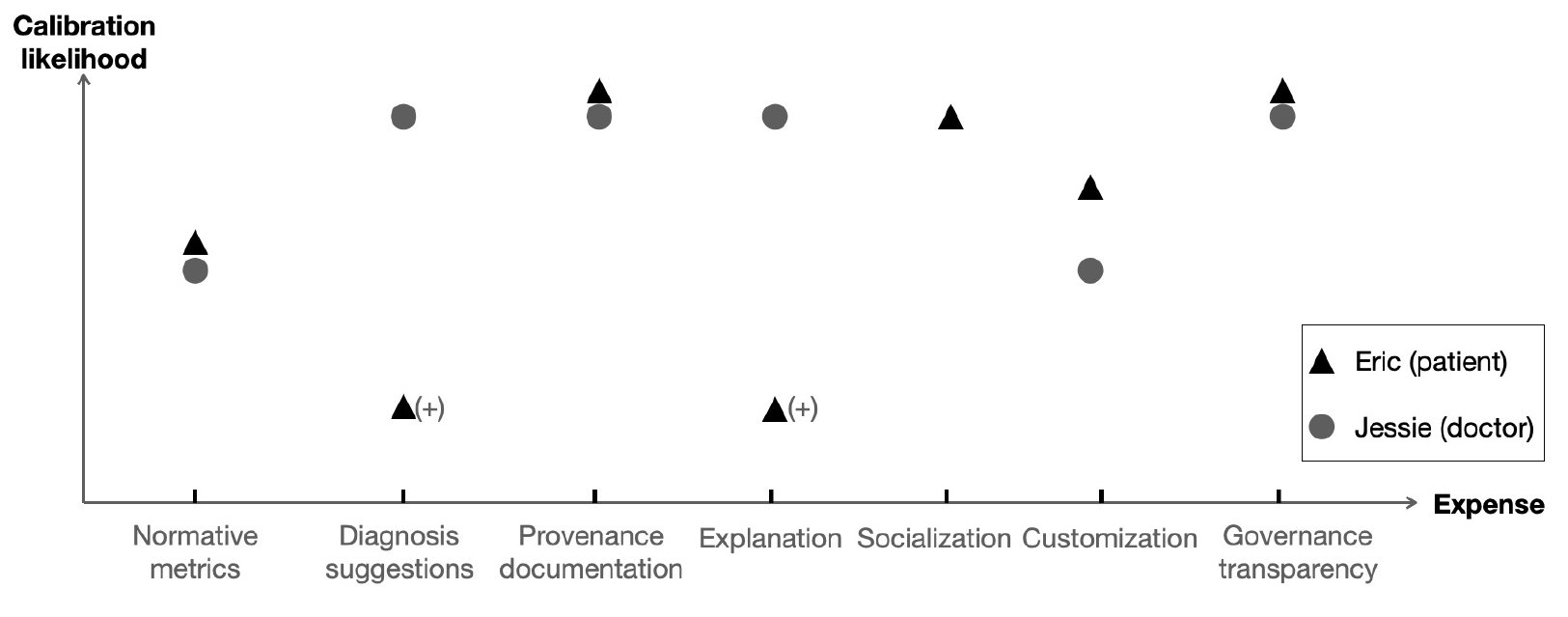}
    \vspace{-1em}
  \caption{Calibration-expense analysis performed on the trustworthiness cues in HealthChecker. $+$ indicates a risk of leading to over-trust.}~\label{fig:analysis}
    \vspace{-2em}
\end{figure*}

Next, they assess the calibration condition for the remaining trustworthiness cues, keeping in mind the two personas, Eric (patient) and Jessie (doctor). They consider the ``systematic condition'' \textit{or} ``no unfounded heuristic condition'' to rate the \textit{calibration likelihood} of each trustworthiness cue. The questions they ask are: 
\vspace{-0.5em}
\begin{enumerate}
    \item Does the user group have the ability or motivation to perform systematic processing? 
    \item If not, what kind of trust heuristic is likely to be invoked?
    \item How likely is this heuristic unfounded and how prominent is the unfounded heuristic?
\end{enumerate}{}
\vspace{-0.5em}
 We recommend answering these questions empirically with target users, e.g., recruiting participants with Eric and Jessica's profiles and conducting think-aloud studies as they interact with the system and make trust judgments. To complete this use case, we survey communication and HCI literature to enlist heuristics that have been identified for relevant trustworthiness cues, as shown in the last box of Figure~\ref{fig:prcess}. We enumerate the analysis for each trustworthiness cue below. The results are summarized in Figure~\ref{fig:analysis}, where the y-axis represents calibration likelihood.

\textit{Diagnosis suggestions}: Jessie is able to make her own diagnosis and reason about the recommendation quality made by the AI. This cue satisfies the systematic condition and has a \textit{high calibration likelihood for Jessie}. Eric lacks the ability to perform a systematic assessment. According to the literature, \textit{machine heuristic} can be prominent for this group of users with a positive attitude towards AI~\cite{sundar2019machine}, which can lead to over-trust. They may also resort to a positive \textit{confirmation heuristic} if the AI's suggestions align with their own speculation~\cite{lu2021human}, which is likely unfounded. Therefore, this cue has a \textit{low calibration likelihood for Eric} and a high risk of leading to over-trust. The creators can improve the design to alleviate the unfounded heuristics, such as presenting uncertainty information and multiple candidate diagnoses.

 \textit{Normative metrics}: Jessie and Eric are not highly proficient with AI metrics, so it is unclear whether these cues can satisfy the systematic condition. Literature suggests that there exists a \textit{numeric heuristic} whereby some people react positively to mathematical information about algorithms~\cite{ehsan2021expanding}. However, there is no reason to believe this heuristic is prominent for Jessie and Eric. The creators consider this cue to have \textit{medium calibration likelihood for both Eric and Jessie}. To improve the calibration, they can provide evidence suggesting the acceptable range of these metrics.

\textit{Explanations}: HealthChecker provides a feature-importance explanation to show how a diagnosis is made based on the most prominent symptoms. Jessie is able to understand and reason about these explanations analytically to assess the AI's process integrity. This cue has a \textit{high calibration likelihood for Jessie}. Recent literature warned against presenting complex explanations to people who lack the ability to understand or assess them but commonly invoke an unfounded \textit{explainability heuristic} that associates being explainable directly with superior capability~\cite{liao2021human}. Hence, explanations may have a \textit{low calibration likelihood for Eric} and can lead to over-trust. The creators should only present explanation designs that are proven to be accessible for users like Eric.
    
\textit{Provenance and governance transparency in documentation}: The creators invested in producing easy-to-read documentation that provides information about model provenance and governance structure. It is also expected that users like Jessie are often motivated to read the documentation for healthcare technologies. So these cues should satisfy the systematic condition and have a \textit{high calibration likelihood for Jessie}. To cater to users unmotivated to spend time reading the documentation, it provides an overview that highlights the source of data used to train the model, the AI principles that the company follows, and certifications from an internal review board. These cues can invoke \textit{source reputation heuristic} and  \textit{authority heuristic}, which are well founded in this context (likely to hold for improving trustworthiness). Hence they satisfy the no unfounded heuristic condition and have \textit{high calibration likelihood for Eric}.
    
 \textit{Customization}:  The customization feature mainly serves Eric to experience the improvability and compliance attributes of the AI. It is reasonable to expect that while using this feature, Eric would invest time to examine the effects although the improvement may not be easy to assess immediately. So the feature has a \textit{medium-high calibration likelihood for Eric}. For Jessie, knowing that the system has a customization function may invoke the \textit{control heuristic} that associates giving user control with a positive intention of the technology creators~\cite{sundar2008main}, which is reasonably founded, but not necessarily a prominent one. So the cue may have a \textit{medium calibration likelihood for Jessie}.
    
\textit{Socialization}: This feature allows Eric to view Jessie's feedback. Positive ratings from Jessie are likely to invoke an \textit{endorsement heuristic} that leads to positive trust judgment. This heuristic is founded in this context and hence this feature provides cues that have a \textit{high calibration likelihood for Eric}. It may not be applicable for Jessie's trust judgment.

\textbf{Step 4: Which ``expensive'' trustworthiness cues should be prioritized?} The creators ask this question to identify trustworthiness cues that would give their product an honest advantage to build user trust and also contribute to good industry practices. They solicit ratings from the team members on the expenses in developing and, if applicable, the processes and infrastructure to obtain, each trustworthiness cue.  The X-axis of Figure~\ref{fig:analysis} represents the results and ranks the more expensive cues to the right. Static presentations of performance metrics and diagnosis suggestions are low-expense because they require only plugging in outputs from the model development pipeline. The explanation visualization requires more effort for UI development. The carefully crafted documentation takes time to produce, and the governance transparency part requires company investment in the infrastructure. The socialization and customization features are expensive for the team and indeed provide competitive advantages over similar products.

\textbf{Summary and guidance for using MATCH.} As illustrated above, practitioners can use MATCH to design AI systems that responsibly communicate their true trustworthiness by following the four-step analysis, after prototypical user groups are identified and properly understood. This calibration-expense analysis, as shown in Figure~\ref{fig:analysis}, is best conducted during the planning stage to help the team identify trustworthiness cues to invest in, by focusing on those that are expensive within their affordable range, and can provide a good calibration for different user groups. It should also be done iteratively as the design and knowledge about target users progress. For example, the creator may attempt to mitigate users' machine heuristic by showing uncertainty information, but an empirical study could reveal that users like Eric have difficulties reasoning about quantitative uncertainty but invoke heuristics that lead to biased interpretation~\cite{hofman2020visualizing,padilla2021uncertain}. In practice, it could be challenging to identify trustworthiness cues and trust heuristics for different user groups exhaustively. We view the current MATCH model as a starting point to engage in careful consideration of the psychological reality of target users, and pinpoint detailed responsibilities in ensuring \textit{appropriate and equitable} user trust for technology creators. We discuss some future directions to advance these practices below.

\section{Discussion: Towards Responsible Trust in AI}
We bring a communication perspective to the discourse on trust in AI. This conceptual work is intended to introduce and synthesize relevant theories on trust in technologies, elucidate the cognitive mechanisms of trust, and call out the requirements for using reliable trustworthiness cues. We invite future research to empirically investigate the topic and develop practical means for building responsible trust in AI, in the following directions.

\textbf{Understanding and regulating the space of trustworthiness cues}. Based on MATCH, technology creators' responsible use of trustworthiness cues has two essential sets of requirements: to truthfully and comprehensively communicate the model trustworthiness attributes, and to use cues based on which the target users are likely to make well-calibrated trust judgments. There are several challenges and complexities for future research to investigate. First, the mapping between cues and trustworthiness attributes is not always one-to-one, meaning that a system feature can cue multiple bases of trust~\cite{kunkel2019let}. It is important to recognize that trust does not reside solely in model ability. It provides an alternative explanation to the observations that adding transparency features often increase people's trust even if the model should not be relied upon~\cite{wang2021explanations,suresh2020misplaced,zhang2020effect,bansal2021does}: they may have enhanced people's intention and process based trust rather than ability based trust. Future research should further unpack the dimensions of trustworthy AI and their relations with conceptually relevant constructs, especially behavioral outcomes such as reliance and compliance~\cite{vereschak2021evaluate}. 

The second challenge arises from our lack of understanding of what constitutes trustworthiness cues in AI systems. A conceptual analysis as we did in this paper is not enough. Future work should empirically study what people actually pay attention to and how they process them when making trust judgments, similar to what has been done in the web trust literature~\cite{fogg2001makes,sillence2004trust,fogg2003prominence,metzger2007making}. To understand the effect of a trustworthiness cue, we echo the point made by ~\citet{jacovi2021formalizing} that it should be studied in relation to different levels of model trustworthiness. This aligns with the common practice in web trust literature where the effect of a web design feature is studied in contrast for web sites with high- versus low-credibility content~\cite{liao2014age}. Such an evaluation protocol allows identifying cues with low calibration (resulting in similar trust judgments for models with different trustworthiness) and also a naturalistic setting to avoid the response bias problem, i.e., users may recognize they should examine certain features yet rarely do so in actual practice~\cite{eysenbach2002empirical,metzger2007making}. Through joint efforts of empirical analysis and theory development, we may outline a more complete design space of reliable trustworthiness cues to guide technology creators' choices~\cite{rieh2007credibility,fogg2001makes}.

\textbf{Empowering users to make accurate trust judgments}. To guard against deceptive or flawed design of trustworthiness cues, a complementary area for responsible trust in AI is to explore means to empower end users to make more accurate trust judgments. Valuable lessons can again be drawn from what researchers have done for supporting web users, among which we highlight two areas of work. One is to provide training materials or guidance for users to assess the system more critically (see review in~\cite{rieh2007credibility}), such as a checklist to assess trustworthiness attributes, and to recognize irrelevant cues or unfounded trust heuristics. The other is to provide independent augmenting tooling to truthfully highlight an AI system's trustworthiness cues, which the creators may have downplayed or hidden~\cite{kittur2008can,schwarz2011augmenting,yamamoto2011enhancing}. \citet{schwarz2011augmenting} developed visualization to augment web search results, displaying metrics that reflect the quality of content in a web site and making visible otherwise hidden information that provides supporting evidence for its level of trustworthiness, such as the web site's PageRank information and visiting patterns of other users. ~\citet{yamamoto2011enhancing} built a system that shows scores of trustworthiness attributes of web sites and re-ranks the search results. 


\textbf{Leveraging model-extrinsic social, organizational, and industrial mechanisms to provide reliable trustworthiness cues}. Communication literature points to many heuristics that people develop through social interactions and based on social structures~\cite{sundar2008main}, and these observations encourage looking into model-extrinsic mechanisms to generate cues that provide supporting evidence for the model trustworthiness attributes. As shown in the example of HealthChecker, a prominent \textit{authority heuristic} can be invoked by communicating the model governance structure; a \textit{source reputation heuristic} can be triggered by communicating the legitimacy of model provenance and track record of service. Also, people have a tendency to follow the opinion of many others (\textit{bandwagon heuristic}~\cite{sundar2008main}). A recent study explored features of ``social transparency'' in AI systems~\cite{ehsan2021expanding}, by showing other users' interaction outcomes and feedback, and found them to help calibrate user trust by tapping into the bandwagon heuristic, among others. When these mechanisms are aligned with efforts needed to establish social, organizational, and regulatory ecosystems for the assurance of trustworthy AI~\cite{shneiderman2020bridging,knowles2021sanction,ehsan2021expanding}, they are likely to satisfy the calibration condition. While there exist non-trivial issues to ensure responsible implementation of these mechanisms and truthful communication, trustworthiness cues from these mechanisms are relatively expensive to obtain, which is another advantage to advocate for their use.

\bibliographystyle{ACM-Reference-Format}
\bibliography{sample-base}
\end{document}